\documentclass[aps,prd,showpacs,nofootinbib,superscriptaddress,preprint,tightenlines]{revtex4}
\newtheorem{definition}{Definition}
\newtheorem{theorem}{Theorem}

\usepackage[latin1]{inputenc}
\usepackage{amsmath}
\usepackage{amsfonts}
\usepackage{amssymb}
\usepackage{pictex}

\def\be{\nopagebreak[3]\begin{equation}}
\def\ee{\end{equation}}
\def\ba{\nopagebreak[3]\begin{eqnarray}}
\def\ea{\end{eqnarray}}

\def\H{{\cal H}}

\def\R{\mathbb{R}}

\newcommand{\teta}{\rlap{\lower2ex\hbox{$\,\tilde{}$}}\eta{}}

\begin{document}
\preprint{\vbox{\baselineskip=12pt \rightline{IGPG-06/10-2}
\rightline{gr-qc/0610072} }}

\title{Hamiltonian and physical Hilbert space in polymer quantum mechanics}
\author{Alejandro Corichi}
\email{corichi@matmor.unam.mx} \affiliation{
Instituto de Matem\'aticas, 
Unidad Morelia, Universidad Nacional Aut\'onoma de
M\'exico, UNAM-Campus Morelia, A. Postal 61-3, Morelia, Michoac\'an 58090, 
Mexico}
\affiliation{Departamento de Gravitaci\'on y Teor\'\i a de Campos, Instituto de Ciencias Nucleares,
 Universidad Nacional Aut\'onoma de M\'exico,\\
A. Postal 70-543, M\'exico D.F. 04510, Mexico}
\affiliation{Institute for Gravitational Physics and Geometry,
Physics Department, Pennsylvania State University, University Park
PA 16802, USA}
\author{Tatjana Vuka{\v s}inac}
\email{tatjana@shi.matmor.unam.mx} \affiliation{Facultad de
Ingenier\'\i a Civil, Universidad Michoacana de San Nicolas de
Hidalgo, Morelia, Michoac\'an, Mexico}
\author{Jos\'e A. Zapata}\email{zapata@matmor.unam.mx}
\affiliation{
Instituto de Matem\'aticas, 
Unidad Morelia, Universidad Nacional Aut\'onoma de
M\'exico, UNAM-Campus Morelia, A. Postal 61-3, Morelia, Michoac\'an 58090, 
Mexico}

\begin{abstract}
In this paper, a version of {\it polymer quantum mechanics}, which is inspired 
by loop quantum gravity, is considered and shown to be equivalent, 
in a precise sense, to the  standard, experimentally
tested, Schr\"odinger quantum mechanics. The kinematical
cornerstone of our framework is the so called polymer
representation of the Heisenberg-Weyl (H-W) algebra, which is the
starting point of the construction. The dynamics is constructed as
a continuum limit of effective theories characterized by a scale,
and requires a renormalization of the inner product. The result is
a physical Hilbert space in which the continuum Hamiltonian can be
represented and that is unitarily equivalent to the Schr\"odinger
representation of quantum mechanics. As a concrete implementation
of our formalism, the simple harmonic oscillator is fully
developed.
\end{abstract}

 \maketitle

\section{Introduction}

A non-standard and `exotic' representation of the canonical
commutation relations, recently introduced in the context of
non-perturbative quantizations of gravity and known as polymer
quantum mechanics has been used to explore both mathematical and
physical issues in background independent theories such as quantum
gravity \cite{AFW}. A  notable example of this type of
quantization, when applied to minisuperspace models has given way
to what is known as loop quantum cosmology \cite{lqc}. As in any
toy model situation, one hopes to learn about the subtle technical
and conceptual issues that are present in full quantum gravity by
means of simple, finite dimensional examples. This formalism
is not an exception in this regard. Such a quantization is made of
several steps. The first one is to build a representation of the
Weyl algebra (sometimes also known as the Heisenberg-Weyl algebra)
on a Kinematical Hilbert space that is ``background independent',
and it is sometimes referred to as the polymeric Hilbert space.
The second part, the dynamics, deals with the implementation of a
Hamiltonian (or Hamiltonian constraint) on this space. In the
examples studied so far, the first part is fairly well understood,
yielding a kinematical Hilbert space $\H_{\rm poly}$ that is,
however, non-separable. Dynamics has proved to be a bit more
difficult, given that a direct implementation of the Hamiltonian
$\hat{H}$ on $\H_{\rm poly}$ is not possible since one of the main
features of this quantum mechanics is that the operators $\hat{q}$
and $\hat{p}$ cannot be both simultaneously defined (nor its analogues 
in theories involving more elaborate variables). Thus, any
operator that involves (powers of) the not defined variable has to
be regulated by a well defined operator. However, the regulator
can not be removed and one is left with the ambiguity that is
present in any regularization. The freedom in choosing it can be
sometimes associated with a length scale, as is the case of a simple harmonic oscillator, where the
freedom can be cast in terms a length scale associated with
the `discreetness of space'. In the standard
treatment of that system, it has been argued that if this length
scale is taken to be `sufficiently small', one can arbitrarily approximate
standard Schr\"odinger quantum mechanics \cite{AFW}\footnote{In loop quantum
cosmology, the minimum area gap of the full quantum gravity theory
imposes such a scale \cite{lqc}.}.

In this paper we shall adopt a different viewpoint. The new
viewpoint is that any scale that one chooses for defining the
Hamiltonian should be seen as providing an effective description
at the given scale. A natural question is what happens when we
change the scale and go to even smaller `distances'. Can we define
consistency conditions between these scales? Or even better, can
we take the limit and find thus a continuum limit? As we shall see
in this paper, the answer to both questions is in the affirmative.

In the remainder of the introduction, let us pose the question
with a bit of more precision. The kinematical Hilbert space
$\H_{{\rm poly},x}$ is the completion of the vector space $Cyl_x$
which is finitely generated by the family of functions $\{
\delta_{x_0} \}_{x_0\in \mathbb{R}}$ (where $\delta_{x_0} (x) = 0$
if $x \neq x_0$ and $\delta_{x_0} (x) = 1$ if $x = x_0$) with the
inner product that makes this family an orthonormal basis. That
is, the ``polymer representation" of the Heisenberg-Weyl algebra
in $\H_{{\rm poly},x}$ \cite{AFW}, is such that the operator
$\hat{x}$ acts by multiplication and the generator of translations
$\hat{V}(\lambda)$ acts as:
$\hat{V}(\lambda)\cdot\Psi(x)=\Psi(x-\lambda)$. The inner product
on the basis is given by:
 $\langle
\delta_{x_0}(x),\delta_{x_1}(x)\rangle=\delta_{x_0,x_1}$. A state
$\Psi \in \H_{{\rm poly},x}$ is of the form $\Psi(x) = \sum_i
\psi_i\, \delta_{x_i} (x)$ with $\sum_i |\psi_i |^2 < \infty$.%
\footnote{This corresponds to the `position representation'
of the polymeric quantum mechanics. In most treatments, the
Hilbert space is described in the `momentum representation' where
the basis is given by quasiperiodic functions and the
inner-product is given by the Haar measure on the Bohr
compactification of $\mathbb{R}$ \cite{lqc}.} 
As we have already
mentioned, the dynamics is  problematic in this framework; not
even the Hamiltonian of the simple harmonic oscillator can be
represented inside $\H_{{\rm poly},x}$. This is the case given
that the operator $\hat{p}$ (the generator of translations in $x$)
is not well defined in $\H_{{\rm poly},x}$, so any observable that
depends on finite powers of $p$ is not well defined as such. In
this regard this simplified quantum mechanical system captures the
essence of loop quantum gravity, including the mathematical origin
of the non triviality of the dynamics. Recall that in the case of
LQG, the curvature of the connection --that appears in the
constraints-- is not a well defined operator so one has to
approximate it by a holonomy.

In this paper we construct a physical Hilbert space and a
Hamiltonian operator on it as a continuum limit of effective
theories. That is, we import the proposal of constructing the
dynamics of loop quantized theories as a continuum limit of
effective theories developed in \cite{MOWZ} to this model
structure for quantum mechanics.

As an example of the continuum limit here presented, the case of
the simple harmonic oscillator is fully developed. We make
extensive use of  previous results by Ashtekar, Fairhurst and
Willis \cite{AFW}. 
However, the viewpoint here advocated is
different from the viewpoint of \cite{AFW}, where the authors
already had evidence of ``some convergence with standard quantum
mechanics" (see also \cite{freden} for a different perspective).
The question that  these papers have tried to address is rather
simple: Can polymer quantum mechanics approximate the standard
representation of quantum mechanics? This question implicitly
comes with the next question: What is the meaning of approximating
a quantum theory? In a sense the difference between \cite{AFW},
\cite{freden} and this work deals with the different answers given
to the later question. The viewpoint here adopted, as we shall
elaborate throughout the paper, is based on the construction of a
continuum theory 
as an appropriately defined limit. 
This theory will turn out to be equivalent to the standard 
Schr\"odinger quantum mechanics. 
Our procedure
involves, as intermediate steps, the construction of effective theories at
different scales, that on themselves can be seen as an
approximation (given certain tolerances) of the continuum theory. 
When we focus attention to one given scale our treatment 
bears some resemblances to the studies presented in \cite{AFW} and
\cite{freden}.

The strategy for constructing the continuum theory starting from
the polymer representation will involve three steps. In the first
one we choose to view the Hamiltonian as a quadratic form, which
implies that it is naturally coarse grained by a map relating
different scales, and that its continuum limit will be a projective
limit. We should explicitly mention that the matching of effective
theories at different scales that is behind the continuum limit
involves coarse graining as well as rescaling. 
The second step
corresponds to working at one given scale,  but replacing the 
effective theory at that scale by better theories 
that know about degrees of freedom from finer scales
and that are generated by coarse graining and rescaling the effective 
theory of a finer scale.  If at a given scale 
these microscopically corrected theories 
converge, the limit is called a completely renormalized theory at that scale. 
In the last step the 
completely renormalized theories at all the scales are pasted together (they satisfy a compatibility condition) to  construct the continuum limit. A product of this construction is the physical Hilbert space. This
space turns out to be equivalent to the ordinary
$L^2$ space of the Schr\"odinger theory.

The structure of the paper is as follows. In Sec.~\ref{sec:2} we
introduce the necessary steps for defining the notion of scale and
the continuum limit at the kinematical level. Section~\ref{sec:3}
is devoted to the dynamics, namely to the Hamiltonian. In this
section the physical Hilbert space is constructed. In
Sec.~\ref{sec:4} we show in detail how the program outlined in the
previous sections is implemented in the case of the harmonic
oscillator. We conclude the paper with a discussion in
Sec.~\ref{sec:5}.

\section{Effective theories, coarse graining and continuum limit: Kinematics}
\label{sec:2}

A central object of the kinematics is the Hilbert space $\H_{{\rm
poly}, x}$; it will be the starting point for our
formalism. In this section we shall see how our notions of
effective theories at a given scale, the coarse graining maps that
relate the different effective theories and the continuum limit
are compatible with $\H_{{\rm poly},x}$.

The role of scales will be played by decompositions of $\R$ as a
disjoint union of closed-open intervals.

\begin{definition}[Scale]
In our context, a scale $C$ is a decomposition of the real line
of the form
\[
\R = \cup_{\alpha_i \in C}\; \alpha_i,
\]
where $\alpha_i= [ L(\alpha_i) , R(\alpha_i) )$, and the vertex
set $\{ L(\alpha_{i+1})= R(\alpha_{i})\}_{\alpha_i \in C}$ is
required not to have any accumulation point.
\end{definition}

Recall that $\H_{{\rm poly},x}$ is the completion of $Cyl_x$, which is finitely generated by the basis given by $\{\delta_{x_0}(x)\}$, consisting of all Kronecker deltas for $x_0\in \R$.
A scale $C$ defines an {\em equivalence relation} $\sim_C$ in
$Cyl_{x}$; we define this relation on one basis and
extend it by linearity to the whole space.  We declare
\[
\delta_{x_0} \sim_C \delta_{y_0} \iff \exists\, \alpha_i \in C :
x_0, y_0 \in \alpha_i\, .
\]

To every scale $C$ we associate a space of {\em states at scale} $C$ 
defined as the
vector space of equivalence classes
\[
Cyl(C) = Cyl_{x} / \sim_C .
\]
This vector space is finitely generated by an orthonormal basis $\{ e_{\alpha}
\}_{\alpha \in C}$ labeled by the cells of $C$.\footnote{There is
a natural projection $P_C: Cyl_{x} \to Cyl(C)$ and the
inner product in $\H_C$ can be pulled back from $\H_{{\rm
poly},x}$ by any section. The end result is that the basis
$\{e_\alpha\}$ becomes orthonormal:
$(e_\alpha,e_\beta)_C=\delta_{\alpha,\beta}$.} The completion of $Cyl(C)$
is the Hilbert space $\H_C$.

We will also work with the dual space $\H_C^\star$. Since the
space of states at any given scale is a Hilbert space, its dual is
isomorphic to it, but for us the dual space will be of special
interest. Notice that its elements have a natural
$\sim_C$-preserving action on $Cyl(\R)_x$. This action is
particularly simple to see for the elements of the dual basis $\{
\omega_{\alpha} \}_{\alpha \in C}$; for them we have
\[
\omega_{\alpha} (\delta_{x_0})= \chi_{\alpha} (x_0)
\]
where $\chi_{\alpha}$ is  the characteristic function of the set 
$\alpha \subset \R$.
Thus, we write
\[
\H_C^\star \subset Cyl(C)^\star
\]
where we have noticed that $Cyl(C)^\star$
can be thought of as the $\sim_C$-preserving subspace of $Cyl(\R)_x^\star$.

\begin{definition}[Coarse graining]
Given two scales we write $C_a \leq C_b$ and say that
$C_a$ is a coarse graining
of  $C_b$ (or $C_b$ is a refinement
of  $C_a$) if any interval $\alpha_i \in C_a$ is a finite union of intervals
of $C_b$.

Our coarse graining maps work by decimation. If we have two scales related
by refinement $C_a\leq C_b$ our decimation map will be defined to be
the injective isometry $d:\H_{C_a} \to \H_{C_b}$ characterized by
\[
d(e_{\alpha})=e_{\beta} \quad \iff L(\alpha) = L(\beta) .
\]
It is important to notice that if $C_a\leq C_b \leq C_c$ the corresponding
$d$-triangle diagram commutes.
\end{definition}

On the other hand, $d^\star:\H_{C_b}^\star \to \H_{C_a}^\star$
sends part of the elements of the dual basis to zero while keeping
the information of the rest: $d^\star (\omega_{\beta})=\omega_{\alpha}$ if $L(\alpha) = L(\beta)$; if there is no
interval $\alpha \in C_a$ with that property $d^\star
(\omega_{\beta})= 0$.

\begin{figure}[htbp]
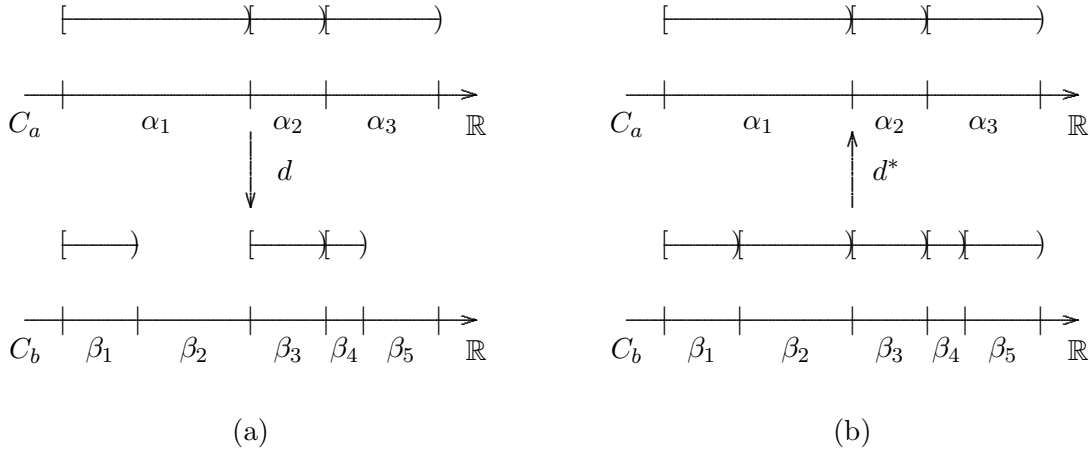

\begin{center}

\input decimacion1.tex

\caption{ (a) $d(\sum_{i=1}^3 e_{\alpha_i})=e_{\beta_1}+e_{\beta_3}+e_{\beta_4}$ \ \
           (b) $d^*(\sum_{i=1}^5 \omega_{\beta_i})=\sum_{j=1}^3\omega_{\alpha_j}$ }
\label{decimation}
\end{center}
\end{figure}

Now we mention two properties that will be central when defining
a continuum limit.

\begin{enumerate}
\item
Apart from being partially ordered, the set of scales is {\em
directed towards refinement}. Any two scales $C_a, C_b$ have a
common refinement, $C_c \geq C_a$ and $C_c \geq C_b$. (In
particular, 
the decomposition consisting of 
the intersection of the intervals of $C_a$ and $C_b$ 
defines a scale that is finer than the original pair of scales.)
\item
The set of scales satisfies the  following property of {\em
infinite refinement}. Given any open set $U \subset \R$ there is a
scale $C$ which has an interval completely contained in it,
$\alpha_i \in C$ and $\alpha_i \subset U$.

\end{enumerate}

Our objective now is to introduce a construction that will let us define the 
limit of $\H_C$ 
as the scale $C$ gets infinitely fine. 
At a given scale $C$ the corresponding Hilbert space $\H_C$ contains ``states 
that come from  coarser scales" $C_0 \leq C$ using the decimation map $d: \H_{C_0} \to \H_C$. It is clear that if $C_0$ 
is not equal to $C$ then there are states in $\H_C$ that do not come from 
states in $\H_{C_0} $. 
Also those states of the form $d (\phi_{C_0})$ may ``come from scales even 
coarser than $C_0$", 
since it could happen that there is a scale $C'_0 \leq C_0$ and a state at 
this scale such that 
$\phi_{C_0} = d (\phi_{C'_0})$. That is, the state $\phi_{C_0} \in \H_{C_0}$ 
and the state 
$\phi_{C'_0} \in \H_{C'_0}$ (such that $\phi_{C_0} = d (\phi_{C'_0})$) agree 
when they are brought to scale $C$ by the corresponding decimation maps. The 
systematic construction that organizes this repeated information is the following. Given a state at any scale coarser than $C$, say $\phi_{C'_0} \in \H_{C'_0}$ with $C'_0 \leq C$, we assign to it a collection 
consisting of one state at every scale $C_0$ that is finer than $C'_0$ 
coarser than $C$, 
$\{ \phi_{C_0} = d (\phi_{C'_0}) \in \H_{C_0} \}_{C'_0 \leq C_0 \leq C}$. In 
addition we must 
regard two such collections as equivalent if they agree at the finer scales; 
that is, the collection 
written above is equivalent to 
$\{ \psi_{C_0} = d (\psi_{C''_0}) \in \H_{C_0} \}_{C''_0 \leq C_0 \leq C}$
if 
and only if 
there is a scale $C_1$ such that $C'_0 \leq C_1$, $C''_0 \leq C_1$ and for 
any scale 
$C_2 \geq C_1$ we have $d (\psi_{C''_0}) = d (\phi_{C'_0}) \in \H_{C_2}$. 
The set of such equivalence classes of collections can be endowed with the 
structure of a vector space and this vector space is isomorphic to $\H_C$ 
itself. The corresponding map, $I_C$ is given by
$$I_C (e_{\alpha_i,C})=\{ (C_0, e_{\beta_j, C_0})|C_0\le C\}$$
where the scale $C_0$ contains an interval $\beta_j$ such that 
$L(e_{\alpha_i,C})=L(e_{\beta_j,C_0})$. Moreover, since the decimation 
maps $d$ are isometries the set of equivalence classes of collections can be 
made a Hilbert space. 

One can notice that the scale $C$ on the collections described above only 
plays the role of 
a cut-off allowing only scales coarser than it. We can work with collections 
that do not have any 
cut-off by simply replacing the scale $C$ by the symbol $\R$. Again we 
can define an 
equivalence relation and it can be given the structure of a Hilbert space. 
This is formalized in the 
following definition.

\begin{definition}[Continuum limit of  $\H_C$]
Our decimation map $d$ gives a meaning to the limit of $\H_C$ as the scale
gets finer and finer. In this case the continuum limit is called a direct limit and
it is denoted by
\[
\stackrel{\longrightarrow}{\H}_{\R} :=
\stackrel{\longrightarrow}{\lim_{C\to \R}} \H_C .
\]
It is defined to be the complete vector
space generated by collections of the type
\[
\phi_{\R} := \{ (C, \phi_C \in \H_C)| C\geq C_0 \; {\rm and} \,\;
\phi_C = d\, \phi_{C_0} \} ,
\]
for some scale $C_0$, and 
where two such collections are identified if they coincide for all
$C \geq C_0'$ for some scale $C_0'$. Addition and multiplication
by scalars is defined ``component by component", and the inner
product is inherited from any of the involved scales because the
decimation maps are isometries.
\footnote{This is exactly the same mathematical structure
appearing in the definition of cylindrical functions. In the
collection written above $C_0$ plays the role of the graph
$\Gamma$ labelling the cylindrical function $\psi_{\Gamma}$. }
\end{definition}

Notice that due to the absence of a cut-off, that allows only certain scales, 
we have to include a clause to guarantee the completeness of the Hilbert space. The effect of such a clause is to add states that have the form of infinite linear combinations of states that ``have their origin at infinitely many distinct given scales". Let us then pose the following result:

\begin{theorem}\label{isomHpoly}
$\stackrel{\longrightarrow}{\H}_{\R}$ is naturally isomorphic to
$\H_{{\rm poly},x}$.
\end{theorem}
To prove this theorem we simply exhibit the isomorphism
$I:\H_{{\rm poly},x} \to \stackrel{\longrightarrow}{\H}_{\R}$,
\[
I( \delta_{x_0} ) = \{
(C, e_{\alpha_{x_0}} \in \H_C)| C\geq C_{x_0}
\} ,
\]
where the decomposition $C_{x_0}$ contains an interval for which
$x_0$ is its left extreme point, and where the interval
$\alpha_{x_0} \in C$ is the one that contains $x_0$.

\begin{definition}[Continuum limit of  $\H_C^\star$]
Also $d^\star$ gives a meaning to the continuum limit of
$\H_C^\star$. Since in this case the limit goes in the direction
opposite to $d^\star$ the limit is called an inverse limit (or
projective limit) and denoted by
\[
\stackrel{\longleftarrow}{\H}_{\R}^\star :=\stackrel{\longleftarrow}{\lim_{C\to \R}} \H_C^\star .
\]
It is defined to be the complete vector space generated by collections
\[
\Phi_{\R} := \{\Phi_C \in \H^\star_C \}
\]
which satisfy the following two properties:
\begin{enumerate}
\item It is  $d^\star$-compatible,
$d_{a,b}^\star(\Phi_{C_b})= \Phi_{C_a}$ whenever $C_a \leq C_b$.
\item The norm of its elements converges, $\lim_{C\to \R} \|\Phi_C\|_C$ exists.
\end{enumerate}
\end{definition}
Notice that the first property guarantees that the elements of
$\stackrel{\longleftarrow}{\H}_{\R}^\star$ have a natural action
on $Cyl(\R)$ given by
\[
\Phi_{\R} (\delta_{x_0}):= \lim_{C\to \R} \Phi_C (\delta_{x_0}) .
\]
The space that we have just described is naturally isomorphic to
$\H_{{\rm poly},x}^\star$. The natural isomorphism $I^\star :
\stackrel{\longleftarrow}{\H}_{\R}^\star \to \H_{{\rm
poly},x}^\star$ is given by
\[
I^\star(\omega_{x_0}) = (\delta_{x_0}, \, \cdot\, )_{\rm poly} ,
\]
where the symbol
$\omega_{x_0}$ is a short hand for the collection
\[
\omega_{x_0} :=\{
\sum_{\alpha \in C}
\delta_{x_0} (L(\alpha)) \omega_{\alpha}
\} .
\]
To summarize we can write
\[
\stackrel{\longleftarrow}{\H}_{\R}^\star \simeq \H_{{\rm
poly},x}^\star \subset Cyl(\R)^\star_x .
\]
This concludes our construction of the kinematical structure
needed for the second step, namely for the definition of dynamics.

\section{Effective theories, coarse graining and continuum limit: Dynamics}
\label{sec:3}

As mentioned in the introduction we will import the proposal of
constructing the dynamics of loop quantized theories as a
continuum limit of effective theories developed in Ref~\cite{MOWZ}
to this model structure for quantum mechanics. 
The basic ideas of this approach are
presented and developed for the quantum field theories in the mentioned article, 
here we give a self-contained treatment for the case of quantum mechanics.
As expected in the
case of quantum mechanics, the procedure is considerably
simplified; only the inner product
will have to be renormalized, but not the coupling constants.

For convenience we will use a smaller auxiliary set of scales
defined by regular decompositions $\{C_n\}$. We start with a regular
decomposition $C_0$ with intervals of length $a_0$, and generate a
totally ordered family of decompositions $\{ C_n \}$ 
from it by regularly
subdividing its intervals repeatedly; then $C_n$'s intervals 
have length $\frac{a_0}{2^n}$. 

The first ingredient is a regularization of the Hamiltonian as a
self-adjoint operator at each scale. With the aid of the inner
product, these Hamiltonians will be treated as quadratic forms;
the quadratic forms that give its expectation value on normalized
states $\lambda_{C_n}^2 ( \psi , h_n   \psi )_{C_n} = h_n(\psi)$
\[
h_n : \H_{C_n} \to \R .
\]
The normalization factors in the inner product $\lambda_{C_n}^2
\in \R^+$ have to be adjusted in such a way that, at least in the
continuum limit, the Hamiltonians of different scales are pasted
correctly by the decimation maps%
\footnote{In standard introductions the continuum limit in the
renormalization group framework includes a wave function
renormalization; here we choose the equivalent action of
renormalizing the inner product instead. We choose to absorb the
normalization factors in the inner product to find a non trivial
action of the completely renormalized Hamiltonian in $\H_{{\rm
poly},x}$ (\ref{convcylstar}).}.

At a given scale $C_m$ we can ``include the effects of more
microscopic degrees of freedom" by using our decimation map. When $C_m \leq C_n$ we define 
\be h_{m(n)} := d_{m,n}^\star h_n .\label{cond-a} \ee
That is, $h_{m(n)}(e_{\alpha_i}) := h_n( d_{m,n}e_{\alpha_i})$. 
If these microscopically corrected Hamiltonians converge, their
limit will be called a completely renormalized Hamiltonian at the
given scale $h_m^{\rm ren} : \H_{C_m} \to \R$,
 \be h_m^{\rm ren} := \lim_{C_n \to \R} h_{m(n)} .\label{cond-0} \ee
By construction, when the completely renormalized Hamiltonians exist, they are
compatible with each other in the sense that 
\[
d_{m,n}^\star h_n^{\rm ren} = h_m^{\rm ren} .
\]
A collection of compatible Hamiltonians defines in itself a
continuum limit Hamiltonian. We will construct a Hilbert space of
physical states after we study the issue of convergence stated in
Eq.~(\ref{cond-0}) in some detail.

We will work with effective Hamiltonians that have a purely
discrete spectrum (labelled by $\nu$) $h_n\cdot\,\Psi_{\nu,
C_n}=E_{\nu, C_n}\,\Psi_{\nu, C_n}$ (if we see the effective
Hamiltonians $h_n$ as quadratic forms the condition is that  they
have isolated critical points $\Psi_{\nu, C_n}$). We shall also
introduce, as an intermediate step, a cut-off in the energy
levels. Thus, we can write 
\be
h_m^{\nu_{\rm cut-off}}  = \sum_{\nu = 0}^{\nu_{\rm cut-off}} E_{\nu, C_m}  \Psi_{\nu, C_m} \otimes
\Psi_{\nu, C_m} \, 
,\label{diag}
\ee
where the eigen covectors $\Psi_{\nu, C_m} \in
\H_{C_m}^\star \subset Cyl(\R)^\star_x$ are normalized according to
the inner product rescaled by $\frac{1}{\lambda_{C_n}^2}$, and the
cut-off can vary up to a scale dependent bound, $\nu_{\rm cut-off}
\leq \nu_{\rm max}(C_m)$. The origin of this cut-off is in the
regularization which 
models the Hamiltonian of our system at a given scale 
with a Hamiltonian of a periodic system in a regime of small
energies. Note that in the polymer representation the operator of momentum $\hat{p}$ is not defined, so it
has to be aproximated using the translation operator, that is well defined. As a
result of this aproximation the standard kinetic term $p^2/2m$ is replaced by a
periodic funcion of $p$.%
\footnote{In the case of the harmonic oscillator we shall
see in detail that the system can be described in terms of a periodic potential, namely, that of a pendulum.}

In  the presence of a cut-off, 
the convergence of the microscopically corrected Hamiltonians,
equation (\ref{cond-0}) is equivalent to the existence of the
following two limits. The first one is the convergence of the
energy levels,
\be \lim_{C_n\rightarrow \R} E_{\nu,C_n} = E_{\nu}^{\rm ren}\,
.\label{cond-1} \ee
Second is the existence of the completely renormalized eigen
covectors,
\be \lim_{C_n\rightarrow \R}
d_{m,n}^\star\,\Psi_{\nu,C_n}=\Psi_{\nu,C_m}^{\rm
ren}\in\H^\star_{C_m}\subset Cyl^\star_x\, .\label{cond-2} \ee
When the completely renormalized eigen covectors exist, they form
a collection that is $d^\star$-compatible.

In addition one could ask whether the eigen covectors
$\Psi_{\nu,C_m}^{\rm ren}$ converge 
in the continuum limit 
as elements of $Cyl(\R)^\star_x$. 
We can choose the normalization factors $\lambda_{C_n}^2$  in a way that that the renormalized eigen covectors converge point-wise
when acting on cylindrical functions $\delta_{x_0}$ when the point
$x_0$ is a vertex of one of the intervals at some scale $C_n$. 
In most cases the dynamics leads to completely renormalized eigen
covectors that are continuous in the sense that they also converge
point-wise when acting on other cylindrical functions. In these
cases, the collection of $d^\star$--compatible Hamiltonians can be
extended to act on $\H_{{\rm poly},x}$. This Hamiltonian (with
cut-off) $h_{\R}^{\nu_{\rm cut-off}{\rm ren}} : \H_{{\rm poly},x}
\to \R$ in the continuum is defined by,
\be h_{\R}^{\nu_{\rm cut-off}{\rm ren}} (\delta_{x_0}) :=\lim_{C_n \to \R} h_n^{\nu_{\rm cut-off}{\rm ren}}
([\delta_{x_0}]_{C_n}) . \label{convcylstar}
\ee

Notice that the domain of this Hamiltonian is $\H_{{\rm poly},x}$ and not only $Cyl_x$; this happens even when the domain of the $\Psi_{\nu,C_m}^{\rm ren}$ is only 
$Cyl_x \varsubsetneq \H_{{\rm poly},x}$ because the quadratic nature of the Hamiltonian improves convergence, so we can extend the action from $Cyl_x$ to
$\H_{{\rm poly},x}$.
From now on, we will assume that  the limit in equation (\ref{convcylstar}) exists for any $x_0 \in \R$. 

This continuum limit Hamiltonian can be coarse grained to {\em
any} scale $d^\star h_{\R}^{\nu_{\rm cut-off}{\rm ren}} : \H_C \to
\R$. Clearly, for a regular decomposition we have $d^\star
h_{\R}^{\nu_{\rm cut-off}{\rm ren}} = h_n^{\nu_{\rm cut-off}{\rm
ren}} $. 

Now we return to the construction of a Hilbert space of physical
states. Recall that an inner product renormalization was used to
obtain our results concerning the continuum limit. Had we 
insisted on using normalized eigen covectors 
$\{\frac{\Psi_{\nu, C_n} }{|| \cdot ||} \}$
according to an inner product which did not have the normalization 
factors, we would have found that the limit in equation (\ref{cond-2}) 
did not exist.  In this case the norm of the eigen covectors diverge 
as $C_n\to \R$.

A choice of normalization factors that leads to convergence in
(\ref{cond-0}) for ``standard systems''%
\footnote{
``Standard systems'' include those with classical Hamiltonian of the form 
$H= p^2/2m + V(x)$. 
} 
is $\lambda_{C_n}^2 = 2^n$, 
which means that the renormalized inner product in $\H_n^\star$ is
\be (\omega_{\alpha_i} , \omega_{\alpha_j})_{C_n}^{\rm ren} = \frac{1}{2^n}
 \delta_{ij} .
\ee
The Hilbert space of covectors together with such inner product
will be called $\H_{C_n}^{\star{\rm ren}}$.

The sequence of compatible renormalized eigen covectors $\{
\Psi_{\nu, C_n}^{\rm ren} \}$ does define an element of
$\stackrel{\longleftarrow}{\H}_{\R}^{\star{\rm ren}}$, which is
the projective limit of the renormalized spaces of covectors \be
\{ \Psi_{\nu, C_n}^{\rm ren} \} \in
\stackrel{\longleftarrow}{\H}_{\R}^{\star{\rm ren}} = \stackrel{\longleftarrow}{\lim_{C_n\to \R}} \H_{C_n}^{\star{\rm ren}}
. \ee
The inner product in this space is defined by
\[
( \{ \Psi_{C_n} \} , \{ \Phi_{C_n}\} )^{\rm ren}_{\R} := \lim_{C_n
\to \R} ( \Psi_{C_n} ,  \Phi_{C_n} )^{\rm ren}_{C_n} .
\]
This inner product is degenerate. Collections of the form
\[
\omega_{x_0} := \{ \sum_{\alpha_i \in C_n} \delta_{x_0}
(L(\alpha_i)) \omega_{\alpha_i} \}
\]
have zero norm, since there is at most one term contributing.
While collections corresponding to characteristic functions of
intervals that fit inside some decomposition $\alpha_i \in C_m$
\[
\tilde{\chi}_{\alpha_i} := \{ 0 \quad \hbox{if} \quad C_n \leq C_m
, \sum_{\alpha_j \in C_n} \chi_{\alpha_i} (L(\alpha_j))
\omega_{\alpha_j}  \quad \hbox{if} \quad C_m \leq C_n \}
\]
now belong to the Hilbert space and their squared norm is the
length of $\alpha_i$ in units of $a_0$.

Given that the inner product identifies many states, the natural
definition of the Hilbert space of physical states is to take the
quotient by the zero norm states, namely:
\begin{definition}[Hilbert space of physical states]
\[
\H_{\rm phys}^\star :=\stackrel{\longleftarrow}{\H}_{\R}^{\star{\rm ren}} / \ker (\cdot
,\cdot )^{\rm ren}_{\R}
\]
\[
\H_{\rm phys} := \H_{\rm phys}^{\star \star}
\]
\end{definition}

\begin{theorem}\label{isom}
The Hilbert space of physical states is naturally unitarily isomorphic to
the space of square integrable functions with respect to the
Lebesgue measure
\[
\H_{\rm phys} \simeq L^2(\R, {\rm d}x)
\]
\end{theorem}
{\em Proof}. 
We will define a unitary anti-isomorphism (invertible antilinear map that preserves the inner product) from 
$L^2(\R, {\rm d}x) \to \H_{\rm phys}^\star$. 

Consider in $L^2(\R, {\rm d}x)$ elements of the form 
$\Psi = \sum_{\alpha} \Psi(\alpha) [\chi_{\alpha}]_{L^2}$, where the sum runs over 
finitely many closed-open intervals and $\chi_{\alpha}$ is the characteristic function of the corresponding interval. The subset of elements of this form is dense in $L^2$.  
Thus, our unitary anti-isomorphism is defined by assigning 
$\Psi = \sum_{\alpha} \bar{\Psi}(\alpha) [\chi_{\alpha}]_{\rm phys}$
to those elements. 
$\Box$

Notice that we could have worked with the set  of all the scales
$C$ (not necessarily corresponding to regular decompositions)
using the corresponding  renormalized inner products for them and
we would have obtained the same result. The use of our auxiliary
set of scales corresponding to regular decompositions simply
selects a dense subset of the physical Hilbert space.



Now we return to the definition of the Hamiltonian in the
continuum limit. 
We can use the renormalized inner product to induce an action of the 
cut--off Hamiltonians 
on $\stackrel{\longleftarrow}{\H}_{\R}^{\star{\rm ren}}$
\[
h_{\R}^{\nu_{\rm cut-off}{\rm ren}} (\{ \Psi_{C_n} \}) := \lim_{C_n \to \R} h_n^{\nu_{\rm cut-off}{\rm ren}} ((  \Psi_{C_n}
 , \cdot )^{\rm ren}_{C_n}) ,
\]
where we have used the fact that $( \Psi_{C_n}, \cdot )^{\rm
ren}_{C_n} \in \H_{C_n}$. 
The existence of this limit is trivial because the renormalized Hamiltonians are finite sums and the limit exists term by term. 

These cut-off Hamiltonians descend to the physical Hilbert space 
\[
h_{\R}^{\nu_{\rm cut-off}{\rm ren}} ([\{ \Psi_{C_n} \}]) := h_{\R}^{\nu_{\rm cut-off}{\rm ren}} (\{ \Psi_{C_n} \}) 
\]
for any representative $\{ \Psi_{C_n} \} \in  [\{ \Psi_{C_n} \}] \in \H_{\rm phys}^\star$.
The context will prevent any confusion with the notation.

Finally we can address the issue of removal of the cut-off. The
Hamiltonian $h_{\R}^{\rm ren} :
\stackrel{\longleftarrow}{\H}_{\R}^{\star{\rm ren}} \to \R$
is defined by the limit
\[
h_{\R}^{\rm ren} := \lim_{\nu_{\rm cut-off} \to \infty}
h_n^{\nu_{\rm cut-off}{\rm ren}}
\]
when the limit exists. 
Its corresponding Hermitian form in $\H_{\rm phys}$ is
defined whenever the above limit exists.

Moreover, we can construct $h_{\R}^{\rm ren} : \stackrel
{\longleftarrow}{\H}_{\R}^{\star{\rm ren}} \to \R$ without
introducing an auxiliary cut-off. We can simply define 
\be
h_{\R}^{\rm ren} (\{ \Psi_{C_n} \}) := \lim_{C_n \to \R} h_n ((  \Psi_{C_n}
 , \cdot )^{\rm ren}_{C_n}) \label{w.o.cut-off} 
\ee 
and arrive to the same
Hamiltonian in the continuum as the previous equation. On the
other hand, the cut-off can not be removed when working on
$\H_{{\rm poly},x}$, or at a given scale; the completely
renormalized Hamiltonians can not be liberated from their cut-off given that,
as we explained earlier, at any given scale the cut-off
is nedeed in order to aproximate the given quantum mechanical system by a periodic one.
There is no solution  to this ``problem'' but, as we will show, it
is really not a problem, but simply an issue related to the domain
of the Hamiltonian.

With the use of the decimation map, the elements of $\H_{C_m}$ have an
action as linear forms on
$\stackrel{\longleftarrow}{\H}_{\R}^{\star{\rm ren}}$
\[
e_{\alpha_i}  ( d_{m,n}^\star (\{\Psi_{C_n} \} ) ) := e_{\alpha_i}
( \Psi_{C_m} ) .
\]
Also, the elements of $\H_{{\rm poly},x}$ act on
$\stackrel{\longleftarrow}{\H}_{\R}^{\star{\rm ren}}$
\[
\delta_{x_0} (\{ \Psi_{C_n} \} ):= \lim_{C_n \to \R}
[\delta_{x_0}]_{C_n} ( \Psi_{C_n} ) .
\]
If any such action is restricted to smooth functions with compact
support, ${\cal C}^\infty_0 \subset
\stackrel{\longleftarrow}{\H}_{\R}^{\star{\rm ren}}$, it becomes
the standard action of Dirac's delta functional.
The natural inclusion of ${\cal C}^\infty_0$ in
$\stackrel{\longleftarrow}{\H}_{\R}^{\star{\rm ren}}$ 
{\em is by an antilinear map} 
which assigns to
any $\Psi \in {\cal C}^\infty_0$ the $d^\star$-compatible
collection 
$\Psi^{\rm shad}_{C_n} := \sum_{\alpha_i} \omega_{\alpha_i} \bar{\Psi}
(L(\alpha_i)) \in \H_{C_n}^{\star{\rm ren}}\subset Cyl^\star_x$; 
$\Psi^{\rm shad}_{C_n}$ will be called the shadow of $\Psi$ at scale $C_n$ and acts in $Cyl_x$ as a piecewise constant function. 
Clearly other types of test functions like
Schwartz functions are also naturally included in
$\stackrel{\longleftarrow}{\H}_{\R}^{\star{\rm ren}}$. This
natural inclusion map is compatible with the isomorphism of
Theorem \ref{isom}.

When $\H_{C_m}$ or $\H_{{\rm poly},x}$ are seen from point of view of
the natural home of the dynamics in the continuum they correspond
to highly distributional objects which are not in the domain of
the Hamiltonians with the standard kinetic term. 
For example, the calculation of equation 
(\ref{convcylstar}) corresponds, in the Schr\"odinger representation, 
to the expectation value of the cut--off Hamiltonian on 
$\delta_{x_0}^{\rm Dirac}$. 

Notice that equation (\ref{w.o.cut-off}) leads to a very compact
version of the dynamics of polymeric quantum mechanics. We can say
that a dynamics in this framework is characterized by a collection
of effective Hamiltonians such that  the limit $h_{\R}^{\rm ren}
$ in
(\ref{w.o.cut-off}) exists for all the elements of a dense domain,
$[\{ \Psi_{C_n} \}]_{\rm phys} \in D_{\rm phys} \subset \H_{\rm
phys}^\star$.

Now we can compare with the standard quantum mechanics resulting
from the Schr\"odinger representation, where the dynamics is taken
to be defined by a Hamiltonian quadratic form $h_{\rm Schr}$
defined on a dense domain $[\Psi]_{L^2} \in D_{L^2} \subset
L^2(\R, {\rm d}x)$. These two dynamics are equivalent if the unitary anti-isomorphism
of Theorem \ref{isom} identifies the domains of definition
and the Hamiltonian quadratic forms. In the next section we will
explicitly see that this is indeed the case for the simple harmonic oscillator.

We have completed the main result of this paper, namely to 
construct a physical Hilbert space which turns out to be isomorphic to that of the
standard Schr\"odinger representation. In the next section, we
shall examine a paradigmatic system that has been the subject of
extensive study, namely the simple harmonic oscillator.

\section{Simple Harmonic Oscillator}
\label{sec:4}

In this section, let us consider the example of a Simple Harmonic Oscillator (SHO) with  parameters $m$ and $\omega$, classically
described by the following Hamiltonian
$$H=\frac{1}{2m}\,p^2+\frac{1}{2}\,m\,\omega^2\, x^2 .$$ 
Recall that from these parameters one can define a length scale
$D=\sqrt{\hbar/m\omega}$. In the standard treatment one uses this
scale to define a complex structure $J$ (and an inner product from
it) that uniquely selects the standard Schr\"odinger
representation. In our case,  since the operator
$$\hat{V}(\mu )=e^{i\mu\hat{p}/\hbar}$$
is not weakly continuous in $\mu$ in the polymeric representation, the
corresponding would-be self-adjoint momentum operator $ \hat{p}$
does not exist. As a starting point we take the suggestion of
Ashtekar, Fairhurst and Willis who define (regulate) $\hat{p}$, at
a given scale $C_n$, as \cite{AFW},
\be
\hat{p}:=\frac{i\hbar\, 2^{n-1}}{a_0}\left[\hat{V}\left(
\frac{a_0}{2^n}\right)-
\hat{V}\left(-\frac{a_0}{2^n}\right)\right]\, .\label{oper-p} \ee
However,  the operator $\hat{p^2}$ that appears in the
Hamiltonian, is then defined not as the square of (\ref{oper-p})
but rather as,
\be \hat{p^2}=\frac{\hbar^2\, 2^{2n}}{a_0^2}\left[2- \hat{V}
\left(\frac{a_0}{2^n}\right)-
\hat{V}\left(-\frac{a_0}{2^n}\right)\right]\, . \ee
As they note this election is not unique and is guided by the
requirement of non-degenerate energy levels (that makes it
preferred compared with $(\hat{p})^2$). Then, the Hamiltonian
$\hat{H}_{C_n}:{\cal H}_{C_n}\to{\cal H}_{C_n}$ at scale $C_n$, is given by,
$$\hat{H}_{C_n}=\frac{\hbar^2}{2m}\left(\frac{2^n}{a_0}\right)^2
\left[
2-\hat{V}\left(\frac{a_0}{2^n}\right)-\hat{V}\left(-\frac{a_0}{2^n}\right)
\right]+\frac{1}{2}\,m\,\omega^2 \,{\hat x}^2$$ where the
operators are represented as
$$\hat{V}\left(\frac{a_0}{2^n}\right)\,e_{\alpha_i,C_n}=e_{\alpha_{i+1},C_n}
\, ,$$
and
$${\hat x}\,e_{\alpha_i,C_n}=L(\alpha_i)\,\,e_{\alpha_i,C_n}\, ,$$
with $L(\alpha_i)=\frac{i\,a_0}{2^n}$. Let us now discuss  what is
being done. The original system is rather symmetric for $x$ and
$p$, being quadratic in both coordinates. This means that one can
alternatively think of $x$ or $p$ as the configuration variable.
Then, what one is doing is to replace the `potential' $p^2$ for a
periodic function, for each $n$ as 
$$p^2\approx 2\left(\frac{2^{n}\,\hbar}{a_0}\right)^2\left[1-\cos{\left(\frac{a_0\,p}{2^n\,\hbar}
\right)}\right].$$ 
This is precisely the
potential of a pendulum with frequency $\omega$ and $n$-dependent
length $\ell_n=\frac{\hbar 2^n}{m\omega a_0}$. That is, we are
approximating, for each scale $C_n$ the SHO by a
pendulum that is longer (and with `stronger gravity' since the effective
gravitational constant is given by $g=\frac{\hbar\omega 2^n}{m a_0}$) as we go to
smaller distances. In the limit we do recover the SHO. 
There is, however, an important difference. From our
knowledge of the pendulum, we know that the quantum system will
have a spectrum for the energy that has two different asymptotic
behaviors, the SHO for low energies and the planar rotor in the
higher end, corresponding to 
oscillating and rotating solutions respectively\footnote{Note
that both types of solutions are, in the phase space, closed.
This is the reason behind the purely discrete spectrum. The
distinction we are making is between those solutions inside the
separatrix, that we call oscillating, and those that are above it that
we call rotating.}. As we refine our scale and both the length of
the pendulum and  the height of the periodic potential increase,
we expect to have an increasing number of oscillating states (for a
given pendulum system, there is only a finite number of such
states). Thus, it is justified to consider the cut-off in the
energy eigenvalues, as discussed in the last section, given that
we only expect a finite number of states of the pendulum to
approximate SHO eigenstates. With these consideration in mind, the
next question is whether the conditions (\ref{cond-1}) and
(\ref{cond-2}) are satisfied for the SHO. 

\begin{figure}[htbp]
\begin{center}
\input{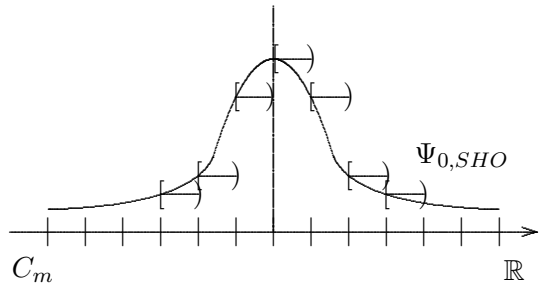}
\caption{{
The solid continuous line represents the  graph of the vacuum state 
$\Psi_{0,{\rm SHO}}$ and the piecewise constant function 
represents $\Psi_{0 ,C_m}^{\rm ren}$.}} \label{convergence}
\end{center}
\end{figure}

We denote the corresponding eigen-covectors as $\Psi_{\nu
,C_n}=\sum_k \bar{\psi}_{\nu
,C_n}\!\!\left(\frac{ka_0}{2^n}\right)\,\omega_{\alpha_k,C_n}
\in{\cal H}_{C_n}^{*ren}$. Let us take into account the
microscopic corrections and define the renormalized
eigen-covectors, at the scale $C_m$ as
$$\Psi_{\nu ,C_m}^{ren}=\lim_{C_n\to\mathbb{R}}d^*_{m,n}\Psi_{\nu,C_n}\ .$$
We can now prove the following result
\be
\Psi_{\nu ,C_m}^{\rm ren}=\Psi^{\rm shad}_{\nu ,C_m}\, 
,\label{reneqshad}
\ee
where $\Psi^{\rm shad}_{\nu ,C_m}(e_{\alpha_i, C_m})=\bar{\psi}^{\rm
Schr}_\nu (ia_0/2^m)$ is (the complex conjugate of) the eigenfunction in the representation of
Schr\"odinger calculated at the extreme left side of the interval
$\alpha_i$ of the partition $C_m$ (see FIG \ref{convergence}). 
The proof is based on the
following numerical result in \cite{AFW}, obtained for the case of
a fixed regular lattice (for our considerations are relevant only
the results referring to $x_0=0$ in their notation) and modified
by using the renormalized inner product,
\be 2^{n/2}\parallel \Psi_{\nu ,C_n}-\Psi^{\rm shad}_{\nu
,C_n}\parallel^{\rm ren} \sim (\nu
+1)^{1.35}\left(\frac{a_0}{2^n D}\right)^{1.10}\, ,\label{desi-1}
\ee
where $D^2=\frac{\hbar}{m\omega}$. This result  is obtained for
$10^{-6}<\frac{a_0}{2^n D}<1$.
Using the properties of $d^\star$ we can easily prove that
$$\parallel d^{\star}_{m,n}\Psi_{\nu ,C_n}-\Psi^{\rm shad}_{\nu ,C_m}
\parallel^{\rm ren} := \parallel d^{\star}_{m,n}(\Psi_{\nu ,C_n}-
\Psi^{\rm shad}_{\nu ,C_n})\parallel^{\rm ren} \leq\ 2^{(n-m)/2}\parallel
\Psi_{\nu ,C_n}-\Psi^{\rm shad}_{\nu ,C_n}\parallel^{\rm ren} \, ,
$$ which, by means of (\ref{desi-1}), leads us to the wanted 
convergence result.
Note that the numerical result of \cite{AFW} is valid in a range
of values that implies a maximum number $n^\prime$. However, there
is strong evidence to assume that the desired result holds valid
(see \cite{AFW} and \cite{aunola} for further analytic arguments).

Note that the $\Psi_{\nu ,C_n}^{\rm ren}$ are compatible in the sense
that
$$d^*\Psi_{\nu ,C_n}^{\rm ren}(e_{\alpha_i,C_{n-1}}) =
\Psi_{\nu ,C_{n-1}}^{\rm ren}
(e_{\alpha_i,C_{n-1}})\ .$$
For {\em any} $\delta_x\in{\cal H}_{{\rm poly},x}$ we have
$$\lim_{C_n\to\mathbb{R}}\Psi_{\nu ,c_n}^{\rm ren}(\delta_x)\equiv
\Psi_{\nu,\mathbb{R}}(\delta_x)=\bar{\psi}_\nu^{\rm Schr}(x)\ ,$$ so
that in this limit we recover all the information about the
eigenstates from the Schr\"odinger quantum mechanics.

Regarding the energy levels we take the corresponding result from
\cite{AFW} and rewrite it in our notation leading to
$$E_{\nu ,C_n}\sim (2\nu +1)\frac{\hbar\omega}{2}+{\cal O}
\left(\frac{a_0}{2^n D}\right)\, ,$$
valid for $(a_0/2^n D)\ll 1$, so that
$$\lim_{C_n\to\mathbb{R}}E_{\nu ,C_n}=E_{\nu ,{\rm SHO}}\, ,$$
where $E_{\nu ,\,{\rm SHO}}$ are the energy levels for the
harmonic oscillator in the Schr\"odinger representation.

The convergence of the eigen covectors and of the discrete energy
levels imply that we have completely renormalized Hamiltonians for
any allowed value of the cut-off
$$h_m^{\nu_{\rm cut-off},{\rm ren}}:=\lim_{C_n\to\mathbb{R}}d^\star_{m,n}h_n^{\nu_{\rm cut-off}} .$$ 
Also the Hamiltonian
$h_{\R}^{\rm ren} : \stackrel
{\longleftarrow}{\H}_{\R}^{\star{\rm ren}} \to \R$ as defined in equation 
(\ref{w.o.cut-off}) (or equivalently by taking the limit $\nu_{\rm cut-off} 
\to \infty$) can be calculated. 
The calculation is especially simple when we consider the states 
$\{ \Psi^{\rm shad}_{\nu ,C_n} \} \in \stackrel
{\longleftarrow}{\H}_{\R}^{\star{\rm ren}}$. After using the convergence result (\ref{reneqshad}) it is easy to find 
\be
h_{\R}^{\rm ren} (\{ \Psi^{\rm shad}_{\nu ,C_n} \}) = E_{\nu ,{\rm SHO}} . 
\ee 
Since the $L^2$-classes corresponding to states of this type form a dense subset of $L^2$, we have shown that the unitary anti-isomorphism of Theorem \ref{isom} identifies our Hamiltonian and 
the Hamiltonian  of the SHO written in the 
Schr\"odinger representation. We
can thus conclude that we have recovered the standard treatment of
the harmonic oscillator when applying our formalism.

Let us end this section with a remark. As we have seen in the system 
considered in this section, there is no intrinsic length scale
that one can consider as fundamental and below which the theory
should not be considered. One could imagine that there is such a
minimum scale, the Planck length, which would provide a cut-off in
the scales $C_n$, namely with a maximum number $N_{\rm Pl}$ for
$n$, but as our results here suggest, we do not need to take such a
cut-off.
Furthermore, even if one chose to have such a cut-off $N_{\rm Pl}$ near the
Planck scale, one might not be able to describe in an accurate
manner, arbitrarily large, macroscopic systems, such as a stone
attached to a coil. The larger the system, the higher the
amplitude of the pendulum potential needed to approximate the
system, so that one could even need a scale $C_n$ well beyond the
Planck scale. How could one then justify the minimum
length in that case? Could one consider, for instance, the 
polymer description as
fundamental, but only for elementary systems, and not for
composite systems such as a large stone? We shall leave these
questions for future research.

\section{Discussion and Conclusions}
\label{sec:5}

Let us summarize our results. We have shown that by starting with
the usual polymer representation of a quantum mechanical system,
we can implement a renormalization procedure that allows us to
define a continuum theory where the physical Hilbert space and the
Hamiltonian are well defined. We proceeded in three steps. The
first one was to consider an equivalent formulation of the polymer
representation, where we could define the notion of scale
refinement and the continuum limit at the kinematical level. In
order to implement a Hamiltonian, that is regarded as a quadratic
form, we showed that, under very mild assumptions, one could
perform a renormalization of the inner product and arrive at a
physical Hilbert space. This space is equivalent to the standard
$L^2$ space of Schr\"odinger quantum mechanics. In order to make
contact with other approaches, we have shown that if certain
conditions are fulfilled, the shadows of the Schr\"odinger
system can be approximated by the continuum polymeric system, in
such a way that in the limit, both systems coincide. We have
analyzed in detail the case of a simple harmonic oscillator, were
all the steps can successfully be completed. The program is
general enough that it can, in principle, be implemented to other
physically motivated systems where the polymer representation is
also the starting point, and to  models defined by a bounding
potential and a discrete spectrum. Work in this direction is
underway, as well as in extending the formalism to systems that in
the standard representation have also a continuum energy spectrum and for
totally constrained systems.

An important issue regarding loop quantization and the
applicability of these methods to different physical systems has
to do with the adequacy of loop quantization for dealing with {\it
all} class of systems. That is, are loop methods tailored to
background independent theories only? And, why should we expect
that these methods are useful for background dependent theories at
all? Our considerations here have been motivated by our
expectations for loop quantization, namely, that it should be
applicable to theories that depend on a metric background as well
as to theories that are independent of such structure. Thus we
expect that, when applied to the case of metric background
dependent theories, it should recover known physics while giving
new theories when applied in the background independent category.
Such a goal can only be obtained if loop quantization itself knows
of no preferred background metric and the dynamics of each
particular theory (that might itself depend on extra structures)
brings in the knowledge of the background structure that it needs.
This is precisely what happens with the polymer representation of
quantum mechanics.

As we have seen, the model of  polymeric quantum mechanics is a
concrete realization of our goal. At the kinematical level we
started with the polymer representation of the Weyl algebra on
$\H_{\rm poly}$. The algebra itself does not know about any extra
structure such as a metric on the configuration space (nor phase
space).
A remarkable fact about $\H_{\rm poly}$ is that it hosts a
faithful unitary representation of the group of homeomorphisms  of
$\R$. 
Later on, we continued with our quantization process and found
that the dynamics called for a renormalization of the inner
product which lead to the physical Hilbert space $\H_{\rm
phys}$ which now knows of a metric structure in $\R$; only
the group of rigid translations can be represented unitarily and
faithfully on it. The moral that this model is trying to teach us
is that, as already mentioned, it is the system under
consideration that introduces extra structure and thus reduces the
symmetries of the final, dynamical theory.

We do not want to claim that our implementation of Wilson's Renormalization 
Group is the only possible one. In fact, the general procedure of \cite{MOWZ} 
(applicable to loop quantized field theories), calls for 
modeling a given scale by a general decomposition of the space 
(or spacetime) manifold. In the case of $\R$, cell decompositions are built out of open intervals ($1$-d cells) and points ($0$-d cells). 
The procedure that we followed here is more economical and it has 
the useful property of being consistent with only one decimation map. A procedure of 
this type is available for $\R^n$ (simply considering the cartesian product of our construction $n$ times) and for $T^n$.  If the procedure is not unique, then 
one must ask if the results can be trusted in any way. 
The existence of the limits defining the continuum limit imply that, as the 
scale is refined, the relevant objects acquire some degree of regularity which 
justifies, to some extent, our initial expectation for robustness of the formalism. 
In cases (like the SHO) where not only the continuum limit exists but 
the relevant wave functions coincide with the shadows of the corresponding 
wave functions on the Schr\"odinger representation 
which in addition happen to be continuous, 
we know that as the scale is refined the wave functions become continuous. In these cases we can expect that other similar procedures would produce the same continuum
limit. 

The particular example we have chosen to analyze, namely the simple harmonic oscillator, has received much attention lately, mainly in connection with a proposed
loop quantization of strings (See for instance \cite{Thomas} and \cite{helling}). 
Even when it might seem that our results contradict in a sense the results of \cite{helling}, this is not the case. This is because we are looking at  different 
physical questions. In our case, we are concerned with a procedure to implement the 
Hamiltonian operator in the continuum, while \cite{helling} analyze a closely related representation where the finite time evolution operator is well defined but neither
$\hat{p}$ nor $\hat{q}$  exist as well defined operators. What the articles in Ref.~\cite{helling} show is that,
by coupling the oscillator to an external field, one might get very different answers for physically motivated quantities. It would be interesting though to apply the procedure here presented to that particular quantization and see whether this behavior
remains in the limit.

What lessons for loop quantum gravity can we learn from our study? 
The first immediate `application' 
of our formalism that comes to mind is loop quantum cosmology. The
classical system possesses a finite number of degrees of freedom,
so it falls within the quantum mechanical category. The relevant operator, namely the
Hamiltonian constraint, does not have the standard form with a kinetic and a potential term, such as a standard particle on a potential, but nevertheless, the methods
of polymer quantization are employed for them \cite{lqc}. 
It is also remarkable that our case study, namely the SHO, has an intrinsic length  scale ($D=\sqrt{\hbar/m\omega}$) that can play a parallel role to that of the 
the Planck length in loop quantum cosmology. 
One could imagine that the intrinsic length scale determines a 
minimum scale, which would provide a `physical cut-off' 
of the type $C_n \leq C_{N_{\rm max}}$, with 
$\frac{a_0}{2^{N_{\rm max}}}= L_{\rm intrinsic}$. 
Our results here are not conclusive for general systems, but they 
suggest that stopping at any given scale might be physically incorrect.

On the other hand, if loop quantum cosmology is to be seen as a
symmetric sector of loop quantum gravity, then one is allowed in
principle to impose such a cut-off, coming from the minimum
area gap of the full theory \cite{lqc}.%
\footnote{Some issues related to the existence of this fundamental
cut-off in LQC have also been raised in \cite{Vel}.} 
 The question remains open,
however, as to whether one should take the Planck scale $C_{N_{\rm
Pl}}$ as fundamental and keep the truncated theory, or to perform
the continuum limit and take the resulting description at face
value. Would in that case the continuum theory coincide with the
Wheeler-DeWitt quantization? or, do the main features of the
truncated theory prevail such as a a bounce near the singularity?
Can one define, for instance, physical observables that implement
the quantization of geometry at the physical Hilbert space level?
Can, therefore, a minimum scale be recovered dynamically?
Work in this direction is underway.

Let us end this paper by asking what might be the relevance of our 
results for the program of loop quantum gravity at large. That is,
can we apply what we have learned  to full loop quantum gravity? 
In this respect, 
we will give one technical comment and later a more general one.  
First, we will elaborate on the issue of ``summing over discrete structures 
versus refining discrete structures" to define the continuum limit (or to get rid of 
the discretization introduced  as a mere auxiliary regularization ingredient). 
Our point of view uses the discrete structures $C$ to define effective 
theories; for example, the states of $\H_C$ are related to equivalence classes of states of $\H_{\rm poly}$. Then when we consider $C' \geq C$, 
the equivalence classes $[ \cdot ]_C$ become unions of 
$[ \cdot ]_{C'}$ equivalence classes. Thus, in our formulation ``refinement"
and ``summing over" can not be separated. Now we turn to comment on a more general issue that calls for attention after our work. A possible
strategy would be to advocate the viewpoint that we should take
seriously the lessons from this simple model as well as from
actual field theories where a more nontrivial Wilsonian
renormalization is needed. Examples of these are 2d Yang-Mills, 3d
quantum gravity and the much more non trivial (scalar) example
that is an interacting relativistic quantum field theory defined
as the continuum limit of 2d Ising field theory \cite{MOWZ}. 
Also, when we apply the formalism of \cite{MOWZ} to gauge theories on a flat metric background, our effective theories and decimation maps lead us to Lattice Gauge Theory, where the renormalization group flow has been extensively studied numerically. 
These studies suggest that one might need to take, in loop quantum
gravity, the Wilsonian viewpoint and perform a nontrivial
renormalization. These and other questions certainly deserve
further investigation.


\section*{Acknowledgments}

\noindent We thank Abhay Ashtekar,  Raymundo Bautista, Homero
D\'\i az,  Robert Helling, Michael Hrusak, Jorma Louko, Seth Major, Gerardo Raggi and  
Jos\'e Velhinho for discussions, suggestions and advise. This work was in part supported by CONACyT
U47857-F and 40035-F grants, by NSF PHY04-56913, by the Eberly
Research Funds of Penn State, by the AMC-FUMEC exchange program and by funds of the CIC-Universidad Michoacana de San Nicol\'as de Hidalgo.

\end{document}